\documentclass{article}
\usepackage{arxiv}

\providecommand{\Description}[2][]{}

\renewcommand{\undertitle}{}
\renewcommand{\headeright}{}

\usepackage[
    backend=biber,
    maxnames=5,
    maxcitenames=2,
    sorting=none,
    style=numeric-comp,
]{biblatex}
\AtBeginBibliography{\small}
\usepackage{common}
\newcommand{\diffcodegen}{\textsc{DiffCodeGen}\xspace}
\newcommand{\passatk}{\textsc{Pass}@$K$\xspace}
\newcommand{\passatone}{\textsc{Pass}@$1$\xspace}
\newcommand{\sstar}{$S^{*}$\xspace}

\newcommand{\Nmodels}{4\xspace}
\newcommand{\pyafl}{\texttt{py-afl}\xspace}
\newcommand{\atheris}{\texttt{Atheris}\xspace}
\newcommand{\llmcodechoice}{\textsc{LLMCodeChoice}\xspace}

\begin{document}

\title{Code Generation by Differential Test Time Scaling}

\author{
    Yifeng He \\
    University of California, Davis \\
    \texttt{yfhe@ucdavis.edu}
    \And
    Ethan Wang \\
    University of California, Davis \\
    \And
    Jicheng Wang \\
    University of California, Davis \\
    \And
    Xuanxin Ouyang \\
    Wuhan University \\
    \And
    Hao Chen \\
    University of Hong Kong \\
    \texttt{chenho@hku.hk}
}
\date{}

\maketitle

\begin{abstract}
	Test-time scaling has emerged as a promising approach for improving code generation by exploring large solution spaces at inference time.
	However, existing methods often rely on public test cases that are unavailable in practice,
	or require extensive LLM inference for candidate selection, leading to significant token consumption and time overhead.
	We present \diffcodegen, a novel test-time scaling method for code generation based on coverage-guided differential analysis.
	\diffcodegen generates diverse code candidates using various sampling and prompting strategies,
	then applies coverage-guided fuzzing to synthesize inputs without requiring any existing tests or large language models.
	By executing all candidates on these inputs, \diffcodegen captures their dynamic behavior and clusters candidates based on behavioral similarity.
	\diffcodegen selects the medoid of the largest cluster as the final output.
	Unlike prior test-time scaling methods that invoke additional LLM inference for candidate selection,
	\diffcodegen performs selection without any extra model calls,
	incurring little to no additional token consumption.
	\diffcodegen is fully asynchronous, naturally suited to the current trend of agentic coding,
	and is thus efficient and highly scalable.
	We evaluate \diffcodegen across \Nmodels large language models,
	demonstrating consistent improvements over baselines.
	Compared to state-of-the-art test-time scaling methods,
	\diffcodegen achieves competitive or superior performance while using only a fraction of time and tokens.
	\diffcodegen is model-agnostic and can be combined with reasoning models to further boost performance.
\end{abstract}


\section{Introduction}

\begin{figure}[t]
      \centering
      \includegraphics[width=.9\linewidth]{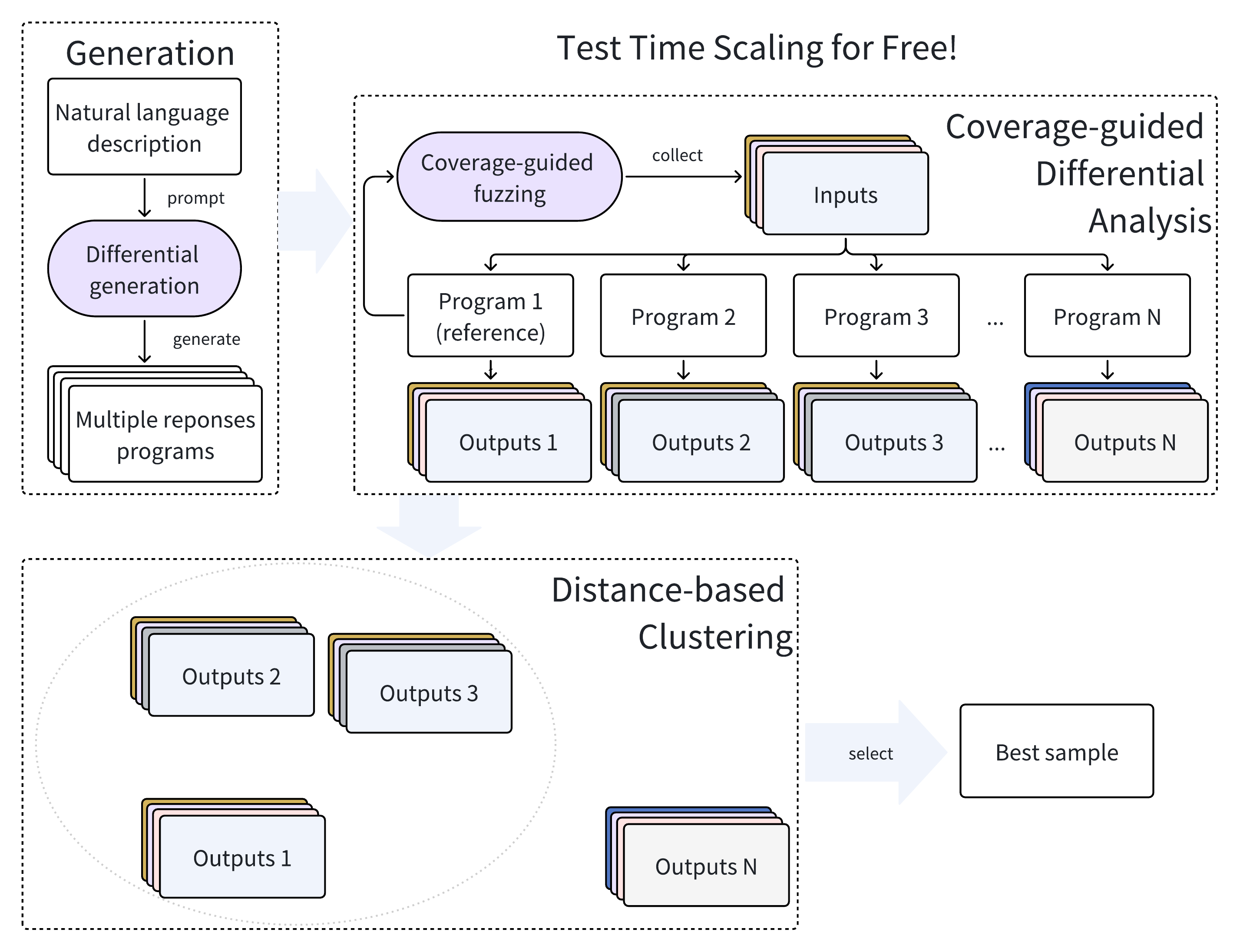}
      \caption{An overview of the \diffcodegen approach. Here, ``for free'' refers to performing candidate selection without any additional LLM inference, incurring no extra token cost beyond the initial candidate generation.}
      \label{fig:diffcodegen}
      \Description{An overview of the \diffcodegen approach.}
\end{figure}

Large language models (LLMs) have become one of the most popular coding assistance tools for software development~\cite{shani2024survey,daigle2024survey}.
Unlike open-ended text generation with natural languages, where quality is often subjective,
code generation has an objective notion of correctness:
the generated program must execute and satisfy the specification.
However, LLM decoding is inherently stochastic.
Sampling creates diversity but also introduces randomness into response correctness,
which often leads to unreliable behavior and unstable performance in practice.

To mitigate this uncertainty during evaluation,
prior work often samples multiple candidates for each coding task in the benchmark.
Let $K$ denote the number of sampled candidates.
Previous work~\cite{chen2021humaneval,du2024evaluating_class_code_gen} reports \passatk,
which considers a task as passed if any of the $K$ candidates passes the tests.
Although \passatk captures the benefit of sampling during evaluation to overcome the possibility that the LLM fails by chance,
it does not match real-world use cases,
where coding assistants must typically return a single solution.
Therefore, selecting a high-quality candidate from a set of samples is crucial for practical code generation.

Test-time scaling (TTS) is a promising direction for searching large solution spaces at inference time~\cite{deepseekai2025deepseekr1,muennighoff2025s1,snell2025scaling,zeng2025acecoder}.
However, many such approaches rely on extensive post-training and reward modeling,
leading to significant resource consumption.
Recent training-free test-time scaling methods for code generation~\cite{li2025sstar}
propose using existing public test cases to iteratively debug generated code,
then use LLM-synthesized inputs to select the best candidate based on dynamic behavior.
Although these methods achieve strong benchmark performance,
their test-available setting limits applicability in practice~\cite{chen2025revisit}.
Furthermore,
these iterative approaches pay the cost of LLM inference twice---once for generating candidates and again for selection---introducing significant time overhead and token consumption.
This motivates the need for a selection method that is both test-free and free of any additional model inference.

Differential testing is a common approach in software engineering to find bugs
across different versions of the same software~\cite{McKeeman1998Differential,evans2007differential_testing,yang2019hunting,li2023finding}.
Intuitively, if most versions behave the same on a set of inputs while a few deviate,
the deviating versions are more likely to be incorrect.
Sampling multiple candidates from a code model naturally fits this setting,
as all of them have the same intended functionality as described by the prompt,
and differences arise only from probabilistic decoding.
Inspired by this observation,
we propose leveraging differential testing to select a candidate from multiple samples without requiring external oracles.

In this work, we introduce \diffcodegen{} (\autoref{fig:diffcodegen}),
a novel test-time scaling method that selects among candidates without invoking any additional LLM inference,
incurring little to no extra token consumption and overhead.
\diffcodegen utilizes different sampling methods to generate diverse candidates from the LLM.
During the selection phase,
\diffcodegen first obtains a set of inputs via coverage-guided fuzzing on one of the candidates,
then executes all candidates on these inputs to collect their outputs to model their dynamic behavior.
Finally, \diffcodegen clusters the candidates based on their dynamic behavior
and selects the candidate with the shortest relative distance to all other candidates in the largest cluster as the final output.
Through these steps,
\diffcodegen effectively selects the most likely correct candidate from a large solution space,
without requiring any extra model inference or token consumption.

\diffcodegen is model-agnostic and can be applied to any LLM.
It can also be combined with reasoning and other test-time scaling methods to further boost performance.
We evaluate \diffcodegen on \Nmodels LLMs,
ranging from small to large and from open-weight to proprietary models.
\diffcodegen consistently improves performance across all models,
and also outperforms existing test-time scaling methods that do not require access to existing public tests.
The main distinguishing feature of \diffcodegen is its low overhead and token cost,
making it a practical choice for real-world code generation tasks.
On \lcb~\cite{jain2025livecodebench},
\diffcodegen takes only one-fifth of the execution time compared to previous TTS methods on locally-served LLMs,
and requires approximately 5\% of the time of previous TTS methods on API-based LLMs.
\diffcodegen also uses significantly fewer tokens compared to existing methods,
consuming only around 4\% of the input tokens used by previous work.
Overall, \diffcodegen presents a highly efficient and effective solution for test-time scaling in code generation.
In summary, our contributions are:
\begin{itemize}
    \item We propose \diffcodegen, a novel test-time scaling method combining differential testing and dynamic software analysis.
          \diffcodegen is both effective and efficient, outperforming existing methods in \passatone performance while using a fraction of time and tokens.
    \item We demonstrate the effectiveness of \diffcodegen across \Nmodels LLMs against two baselines,
          comprehensively evaluating the performance gain of applying \diffcodegen.
    \item We propose and evaluate three distinct differential generation methods, with detailed design and performance analysis of their different characteristics.
    \item We open-source our implementation of \diffcodegen to facilitate future research on applying software testing and analysis techniques to help AI for software engineering.
\end{itemize}

\section{Background}

\subsection{Coverage-guided differential testing}

\subsubsection{Differential testing}
Differential testing is a widely-adopted software testing technique that addresses the problem of \emph{missing oracle},
\ie, the challenge of determining whether a program's output is correct when no ground truth specification exists~\cite{McKeeman1998Differential}.
Instead of relying on expert-provided oracles (\ie, expected outputs),
differential testing compares the behaviors of multiple implementations that should conform to the same specification.
When these implementations produce different outputs on the same input,
at least one of them must contain a bug.

Cross-referencing multiple implementations of the same specification provides a pseudo-oracle,
where discrepancies in outputs indicate potential faults,
and agreement suggests the likelihood of correctness~\cite{McKeeman1998Differential,evans2007differential_testing}.
This technique has proven highly effective for testing compilers~\cite{yang2011finding,le2014emi,sun2016emi_oopsla},
database systems~\cite{rigger2020sqlancer},
and other systems where formal specifications or programmer-provided test oracles are either unavailable or impractical to use.

LLM-based code generation naturally fits the differential testing paradigm.
When sampling multiple code candidates from a language model given the same natural language specification,
each candidate represents an independent attempt to implement the described functionality.
Although these candidates may differ in implementation details,
variable names, control flow structures, or algorithmic choices,
they share the same intended cognitive semantics as defined by the prompt~\cite{he2025tfbench}.
This setting mirrors the classical differential testing scenario:
multiple implementations of the same specification that should produce identical outputs on valid inputs.
Candidates that exhibit consistent behavior across a diverse set of inputs are more likely to correctly implement the specification,
while those that deviate from the majority suggest potential bugs or misinterpretations of the requirements.
By leveraging this insight,
differential testing can serve as an effective oracle-free selection mechanism for LLM-generated code.

\subsubsection{Coverage-guided fuzzing} \label{sec:bg-cov-fuzzing}
Coverage-guided fuzzing uses code coverage as feedback to guide input generation toward unexplored program regions~\cite{bohme2016coverage,fioraldi2020afl_pp}.
Fuzzers instrument the program under test to record which code paths each input exercises,
then prioritize inputs that trigger new coverage~\cite{Zeller2019fuzzingbook}.
This design enables the fuzzer to record the inputs that are meaningful for exploring program behaviors.

A fundamental challenge in differential testing is generating inputs that effectively explore the behavioral differences among implementations.
Random input generation often produces trivial test cases that exercise only common code paths,
missing bugs in the deeply nested branches~\cite{yang2019hunting,li2023finding,chen2019matryoshka}.
To address this limitation,
the combination of differential testing with coverage-guided fuzzing has been proposed to enhance input generation~\cite{chen2016jvm,petsios2017difuzz,nilizadeh2019diffuzz}.
Coverage guidance ensures that the generated inputs explore diverse program behaviors,
while differential comparison across implementations reveals semantic inconsistencies.

This combined approach addresses a key limitation of traditional differential testing:
without coverage guidance,
differential testing may repeatedly exercise the same code paths while missing bugs in less-frequently-executed regions.
Coverage-guided differential testing systematically explores the behavioral space of all implementations under test,
increasing the likelihood of detecting semantic differences that indicate bugs~\cite{chen2019deepdiff}.
For LLM-based code generation,
coverage-guided fuzzing offers a principled alternative to LLM-synthesized test inputs.
While previous work has explored using LLMs to generate test cases for candidate selection~\cite{chen2023codet,li2025sstar},
such approaches inherit a fundamental limitation:
LLMs predict tokens based on statistical patterns in training data rather than executing or simulating program behavior~\cite{he2025tfbench}.
As a result, LLM-generated inputs may fail to expose subtle behavioral differences among candidates.
In contrast, coverage-guided fuzzing directly instruments and executes the generated code,
systematically discovering inputs that explore diverse program states~\cite{he2025fuzzaug}.
This execution-based approach captures the true dynamic behavior of code candidates,
providing a more reliable foundation for differential analysis than purely generative methods.

\subsection{Probabilistic sampling} \label{sec:prob-sampling}
The nature of generative language models is inherently probabilistic.
In language modeling, the probability of generating a given sequence of tokens $t_1, t_2, \ldots, t_n$ is defined as
\begin{equation} \label{eqn:seq-prob}
    P(t_1, t_2, \ldots, t_n) = P(t_1) P(t_2 | t_1) \cdots P(t_n | t_1, \ldots, t_{n-1}) = \prod_{i=1}^{n} P(t_i | t_1, t_2, \ldots, t_{i-1}).
\end{equation}
At each decoding step given all the previous tokens,
generative language models predict a conditional probability distribution over the vocabulary to select the next token
$P(t_i | t_1, t_2, \ldots, t_{i-1})$.
Plain greedy decoding always selects the token with the highest probability,
which often leads to repetitive and low-quality outputs~\cite{holtzman2018learning,holtzman2020curious}.
Instead of always selecting the text that maximizes likelihood,
various probabilistic sampling methods have been proposed to introduce diversity to text generation~\cite{fan2018hierarchical,radford2019gpt2,holtzman2020curious,zhang2025active}.

\subsubsection{Temperature}
The temperature-based sampling parameter is one of the most widely adopted approaches to this problem~\cite{ackley1985learning,hinton2015distilling,ficler2017controlling,fan2018hierarchical,renze2024effect}.
Let $T$ denote the temperature parameter and $t_i$ denote the next token to be generated,
temperature sampling adjusts the conditional probability distribution by scaling the logits of the token $z(t_i)$ before applying the softmax function:
\[
    P_T(t_i | t_1, t_2, \ldots, t_{i-1}) = \frac{\exp\left(\nicefrac{z(t_i)}{T}\right)}{\sum_{j} \exp\left(\nicefrac{z(t_j)}{T}\right)}.
\]
Therefore,
a low temperature sharpens the probability distribution,
thus making the model more deterministic.
In contrast,
a high temperature flattens the probability distribution,
increasing the chance of selecting less likely tokens.
For problem-solving, especially programming tasks,
temperature is often seen as a trade-off between exploration and exploitation:
a high temperature explores the solution space more broadly~\cite{renze2024effect},
but also with the higher chance of generating completely wrong answers~\cite{li2025sstar,xia2025demystifying}.

\subsubsection{Beam search}
Instead of completely maximizing likelihood or reshaping the probability distribution,
beam search maintains a set of highest-probability sequences at each time step~\cite{zhang2023dive}.
Let $k$ denote the beam width, \ie, the number of highly-likely sequences to maintain at each decoding step.
We have a beam set $B_i$ at step $i$, $|B_i| = k$.
Intuitively, we select the top-$k$ tokens at the first decoding step,
each of which is the start of a beam.
At each subsequent decoding step $i$, for each beam sequence in $B_i$,
we append their corresponding top-$k$ tokens to form new candidate sequences, resulting in $k \times k$ candidates.
We select the top-$k$ sequences with the overall highest likelihood from all the candidates to form the new beam set $B_{i+1}$.
Formally, the update step is
\[
    B_{i+1} = \text{Top}_{k} \left\{ P(s)P(v | s): s \in B_i , v \in V \right\},
\]
where $P(s)$ is the likelihood of a token sequence given by \autoref{eqn:seq-prob}
and $v$ is the text token to generate.
At the final generation step,
the beam with the highest overall likelihood is selected as the output.
By its nature, beam search only adds a constant factor of $k$ to the time complexity of decoding
(some view greedy decoding as a special case of beam search with $k=1$),
making it an efficient way to obtain multiple responses from LLMs in a single decoding pass~\cite{zhang2023dive}.

\subsection{On the effects of prompting}
LLMs are brittle to prompt variations~\cite{liang2023holistic,sclar2024quantifying,zhuo2024prosa,errica2025wrong}.
\textcite{sclar2024quantifying} finds that small formatting variations in the prompt
lead to significant performance differences in natural language question-answering tasks across different models.
\textcite{tony2025prompting} finds that different prompting techniques affect the security level of code generated by LLMs on
security-relevant programming tasks.
Variation in prompting is also shown to have a vast effect on the quality of LLM-generated code~\cite{nguyen2024beginning}.
The prompting style of text-to-code tasks is not the only factor,
\textcite{liu2026cream} show that even small semantic-preserving perturbations to prompts
can lead to differences in the generated code's functionality.

\section{Design of \diffcodegen} \label{sec:method}

\subsection{Overview}

In this section, we describe the design of \diffcodegen.
\diffcodegen fits a zero-shot setting in which the user provides a natural language prompt
describing the expected behavior of the resulting program, and the LLM generates code in one response accordingly.
We consider a real-world scenario in which no test suite or example inputs exist for the desired new code,
as public tests rarely exist before a function is implemented.
The workflow of \diffcodegen is illustrated in \autoref{fig:diffcodegen} with three major components:
\begin{enumerate}
      \item \textbf{Differential generation} (\autoref{sec:differential_generation}):
            \diffcodegen generates multiple code candidates from the same prompt
            via different generation strategies.
            We provide and analyze three different approaches in our study.

      \item \textbf{Coverage-guided differential analysis} (\autoref{sec:cov-diff-analysis}):
            \diffcodegen instruments and fuzzes a selected reference solution to collect valid inputs.
            With the inputs,
            \diffcodegen executes all generated code candidates and records their dynamic behavior.

      \item \textbf{Response selection by distance-based clustering} (\autoref{sec:selection}):
            \diffcodegen clusters the generated code candidates based on their dynamic behavior into two clusters.
            Our hypothesis is that the larger one with behavior consensus
            is more likely to contain correct solutions.
            \diffcodegen selects the center of the larger cluster as its final output.
\end{enumerate}

\subsection{Differential generation} \label{sec:differential_generation}

Many approaches exist for obtaining diverse responses from LLMs with the same prompt.
Over the years,
researchers have developed various stochastic sampling, decoding,
and prompting strategies to diversify the outputs of LLMs,
aiming to improve the quality of generated content~\cite{holtzman2020curious,renze2024effect,liu2026cream}.
In \diffcodegen,
we integrate and investigate three different approaches to generate diverse code candidates from the same prompt:
stochastic sampling (with temperature and nucleus sampling), beam-search decoding, and prompt perturbations.

\subsubsection{Differential sampling} \label{sec:diff-sampling}

The core principle behind differential sampling is to exploit the stochastic nature of LLM decoding (\autoref{sec:prob-sampling}) to generate semantically diverse candidates from the same prompt.
For code generation tasks,
the same natural language prompt describes the semantics of the desired program.
No matter how sampling diversifies the implementation details,
all candidates should produce the same execution behavior on valid inputs.
In \diffcodegen,
we control the randomness with both temperature and nucleus sampling to generate diverse code candidates.
We follow previous work~\cite{shi2024can,li2025sstar} to set temperature to $0.7$
and use nucleus sampling with $p=0.95$.
We repeat the sampling process $n$ times for each input prompt to obtain $n$ code candidates.

\subsubsection{Differential decoding} \label{sec:diff-decoding}

Beam search~\cite{freitag2017beam} is a best-first search decoding strategy.
Different from \autoref{sec:diff-sampling} where we repeat the sampling process $n$ times to obtain $n$ diverse outputs,
beam search keeps $n$ best partial sequences at each decoding step,
naturally producing $n$ complete sequences at the end of decoding in a single pass.
At the end of the decoding forward pass,
instead of choosing the single best sequence as the output,
we use all $n$ sequences as our code candidates.

\subsubsection{Differential prompting} \label{sec:diff-prompting}

LLMs are highly sensitive to prompt variations~\cite{sclar2024quantifying,zhuo2024prosa,errica2025wrong}.
Some may view this as a weakness or drawback of generative models and try to avoid it~\cite{jain2025livecodebench},
but in \diffcodegen, we view it as a feature to elevate the chance of generating a better solution.
Previous work~\cite{dhole2023nl} also used small changes in the natural language instructions
as augmentation techniques to improve the performances of NLP models.
To enable differential generation via prompting,
we designed 18 semantic-preserving prompt perturbations to introduce variations to the original prompt.
As shown in \autoref{tab:prompt_perturbations},
none of the perturbations change the core semantics of the original prompt.
With these perturbations,
we generate 18 code candidates from the same prompt by applying each perturbation once.

\begin{table*}[t]
      \centering
      \caption{Summary of 18 prompt perturbation methods used for differential testing, organized by category.}
      \label{tab:prompt_perturbations}
      \begin{adjustbox}{max width=\textwidth}
            \begin{tabular}{@{}lll@{}}
    \toprule
    \textbf{Category} & \textbf{Name}           & \textbf{Description}                                                          \\
    \midrule
    \multirow{4}{*}{\textbf{Lexical}~\cite{dhole2023nl,qiang2024prompt}}
                      & Synonym Replacement     & Replace common words with synonyms (e.g., ``write''$\to$``compose'')          \\[0.5ex]
                      & Formalize               & Convert to formal tone (e.g., ``you''$\to$``one'', ``please''$\to$``kindly'') \\[0.5ex]
                      & Informalize             & Convert to casual tone with ``Hey,'' and ``You gotta nail it!''               \\[0.5ex]
                      & Shorten                 & Remove common adjectives (e.g., ``efficient'', ``optimized'')                 \\
    \midrule
    \multirow{2}{*}{\textbf{Structural}~\cite{zhou2024can}}
                      & Reverse Sentences       & Reverse the order of sentences in the prompt                                  \\[0.5ex]
                      & Shuffle Sentences       & Randomly shuffle the order of sentences                                       \\
    \midrule
    \multirow{4}{*}{\textbf{Informational}~\cite{qiang2024prompt}}
                      & Paraphrase              & Prepend ``In other words,'' to reword the prompt                              \\[0.5ex]
                      & Add Constraint          & Append efficiency and edge case handling requirements                         \\[0.5ex]
                      & Expand Details          & Append scalability and documentation requirements                             \\[0.5ex]
                      & Add Ambiguity           & Append text suggesting multiple interpretations                               \\
    \midrule
    \multirow{3}{*}{\textbf{Stylistic}}
                      & Add Greeting            & Prepend a random greeting (e.g., ``Hello!'', ``Hi there!'')                   \\[0.5ex]
                      & Add Farewell            & Append a random farewell (e.g., ``Cheers!'', ``Best regards!'')               \\[0.5ex]
                      & Add Humor               & Append a humorous warning or note                                             \\
    \midrule
    \multirow{3}{*}{\textbf{Noise}~\cite{pandia2021sorting,fang2024enhancing}}
                      & Add Irrelevant Prefix   & Prepend an unrelated sentence (e.g., about weather)                           \\[0.5ex]
                      & Add Irrelevant Suffix   & Append an unrelated sentence (e.g., ``I love ice cream'')                     \\[0.5ex]
                      & Add Distracting Context & Append distracting background information                                     \\
    \midrule
    \textbf{Error}~\cite{rychalska2019models}
                      & Introduce Errors        & Remove punctuation (commas, semicolons) to simulate typos                     \\
    \bottomrule
\end{tabular}

      \end{adjustbox}
\end{table*}

\subsection{Fuzzing for dynamic behavior modeling} \label{sec:cov-diff-analysis}

After obtaining $N$ solution candidates from differential generation,
\diffcodegen applies \emph{coverage-guided differential fuzzing} to collect their dynamic behavior.

\subsubsection{Coverage-guided fuzzing} \label{sec:fuzzing}
Using coverage-guided fuzzing as described in \autoref{sec:bg-cov-fuzzing},
\diffcodegen collects inputs that explore the dynamic behavior of generated code candidates.
We first select one of the generated code candidates as the \emph{reference solution} to be fuzzed.
For repeated stochastic sampling, we randomly pick one of the candidates;
for beam search, we select the top-1 beam as the reference;
and for prompt perturbations, we select the candidate generated from the original prompt without changes.

In real-world code generation scenarios,
users may instruct the model to implement stand-alone executable programs or library functions.
To accommodate both scenarios,
\diffcodegen supports two modes: \emph{script-based} and \emph{library} fuzzing.
Script-based programs are command-line applications that read input from standard input,
processing the input as a byte stream.
In script-based fuzzing,
the reference solution is treated as a complete program that can be executed directly.
The fuzzer generates inputs that are fed to the program via standard input.
\diffcodegen utilizes \aflpp~\cite{fioraldi2020afl_pp} for script-based programs.
For Python programs, we follow \textcite{zhao2023understanding} to use \pyafl\footnote{\url{https://github.com/jwilk/python-afl}}.
\diffcodegen uses the saved seed queue from \aflpp for the next step.

Library functions, on the other hand, are not directly executable
and require specific function arguments as inputs.
Therefore, fuzzer drivers are used for fuzzing library functions.
These fuzz drivers take random bytes from the fuzzer,
subsequently convert these bytes into well-structured arguments for API calls,
and use the parsed arguments to execute the function under test.
\diffcodegen uses \libfuzzer~\cite{serebryany2016libfuzzer} as the engine for library function fuzzing.
For Python functions,
we use \atheris\footnote{\url{https://github.com/google/atheris}} as the fuzzing framework.
However, unlike script-based fuzzing where we can directly execute the program,
fuzz drivers do not exist for newly LLM-generated library functions.
To this end,
\diffcodegen utilizes an LLM-based approach to generate the fuzz driver,
which is a simplified version of \promptfuzz~\cite{lyu2024promptfuzz}.
Specifically,
\diffcodegen packages the generated library function, the instruction to generate the fuzz driver,
and the documentation of how to parse inputs for the function as a prompt to an LLM.
\diffcodegen generates the fuzz driver with a single pass.
If the generated driver fails to compile or run,
\diffcodegen uses the selected reference solution as the final output.
We follow \fuzzaug~\cite{he2025fuzzaug} to instrument the entry block of functions under test
to record inputs that trigger new code coverage during fuzzing.

\begin{definition}[Program dynamic behavior] \label{def:dynamic_behavior}
      Given a program $p$ and an input set $I_p = \{i_1, i_2, \ldots, i_m\}$,
      the dynamic behavior of $p$ on $I_p$ is defined as a set of 3-tuples
      \[
            B_p = \{(in_i, out_i, err_i) | in_i \in I_p, (out_i, err_i) := \text{Exec}(p, in_i)\},
      \]
      where $out_i$ and $err_i$ are the output and error code of executing $p$ with input $in_i$, respectively~\cite{zhao2023understanding}.
\end{definition}

\subsubsection{Differential analysis}
After fuzzing, we obtain a set of inputs that explore the dynamic behavior of the reference solution.
\diffcodegen then executes all generated code candidates with the collected inputs.
For all code candidates generated from the same prompt,
\diffcodegen ensures the same set of inputs is used for execution.
We collect the outputs, errors, and exit codes of each execution as dynamic behavior of the program as described in \autoref{def:dynamic_behavior}.
For script-based programs,
we capture the standard output and standard error streams;
for library functions, we directly record the returned values and capture any exceptions raised during execution.
For a given programming task, \diffcodegen obtains $\{ (p_i, B_{p_i}) | i \in 1..N \}$ to choose a final solution,
where $p_i$ is the $i$-th generated code candidate.

\subsection{Response selection by clustering} \label{sec:selection}

Instead of relying on an LLM to pick the best solution as in previous work~\cite{li2022alphacode,wang2023selfconsistency,li2025sstar},
which is expensive and inefficient,
\diffcodegen selects its final output by distance-based clustering on the dynamic behavior of all generated code candidates.
\diffcodegen adopts hierarchical agglomerative clustering with average-linkage~\cite{cohenaddad2019hierarchical,tokuda2022revisiting}
to discover behavior consensus among generated code candidates.
The overview of \diffcodegen's clustering-based selection is illustrated in \autoref{alg:clustering},
and we describe the details in the following paragraphs.

\begin{algorithm}[t]
    \caption{Response Selection by Distance-Based Clustering}
    \label{alg:clustering}
    
    \begin{algorithmic}[1]
        \Require Code candidates $P = \{p_1, p_2, \ldots, p_N\}$, dynamic behaviors $\{B_{p_i}\}_{i=1}^{N}$
        \If{$N = 1$}
        \State \Return $p_1$
        \EndIf
        \For{$i \gets 1$ to $N$}  \Comment{Compute pairwise distance matrix $D$ (\autoref{sec:clustering_algorithm})}
        \For{$j \gets i+1$ to $N$}
        \State $I_{ij} \gets I_{p_i} \cap I_{p_j}$ \Comment{Common valid inputs}
        \If{$I_{ij} = \emptyset$}
        \State $D_{ij} \gets 1.0$
        \Else
        \State $D_{ij} \gets d(p_i, p_j)$ \Comment{Dynamic behavior distance \autoref{eq:distance}}
        \EndIf
        \State $D_{ji} \gets D_{ij}$
        \EndFor
        \EndFor
        \State $\{C_0, C_1\} \gets \textsc{AgglomerativeClustering}(D, k{=}2, \text{linkage}{=}\text{average})$
        \State $C^* \gets \argmax_{C \in \{C_0, C_1\}} |C|$ \Comment{Select consensus cluster}
        \State $p^* \gets \argmin_{p_i \in C^*} \sum_{p_j \in C^*, j \neq i} D_{ij}$ \Comment{Find medoid in consensus cluster (\autoref{eq:medoid})}
        \State \Return $p^*$
    \end{algorithmic}
\end{algorithm}

\subsubsection{Normalization of dynamic behavior}
Before comparing dynamic behaviors across code candidates,
\diffcodegen normalizes execution outputs to account for superficial differences that do not affect semantic correctness.
The normalization process handles several common variations:
\begin{enumerate*}
      \item line ending differences (\texttt{\textbackslash r\textbackslash n} and \texttt{\textbackslash r} are converted to \texttt{\textbackslash n});
      \item trailing whitespace is stripped from outputs;
      \item non-string outputs are converted to their string representations.
\end{enumerate*}
This normalization ensures that two programs producing semantically equivalent outputs
are not incorrectly classified as different due to platform-specific or formatting variations.

When comparing two execution results,
\diffcodegen first checks whether the exit codes match.
If the exit codes differ, the results are considered different regardless of their outputs.
For matching exit codes, \diffcodegen compares the normalized standard outputs.
Standard error streams are selectively included in the comparison:
for successful executions (exit code 0), stderr is typically ignored since warning messages may vary across implementations;
for failed executions, stderr is included to distinguish between different failure modes.

\subsubsection{Distance metric}
To quantify the behavioral similarity between two code candidates,
\diffcodegen defines a distance metric based on their dynamic behavior.
Given two programs $p_i$ and $p_j$ with their dynamic behaviors $B_{p_i}$ and $B_{p_j}$ collected from differential analysis,
the distance between them is defined as the fraction of inputs on which they produce different outputs:
\begin{equation} \label{eq:distance}
      d(p_i, p_j) = \frac{|\{in \in I_{ij} \mid \text{Exec}(p_i, in) \neq \text{Exec}(p_j, in)\}|}{|I_{ij}|}
\end{equation}
where $I_{ij} = I_{p_i} \cap I_{p_j}$ is the set of common inputs
on which both programs produce valid execution results,
and $I_{p}$ denotes the input set of program $p$ as defined in \autoref{def:dynamic_behavior}.
Two execution results are considered equivalent if they have the same exit code and produce identical standard output after normalization.
If no common inputs exist between two programs (\ie, $I_{ij} = \emptyset$),
\diffcodegen assigns a maximum distance of $1.0$.
This pairwise comparison strategy ensures that a single crashing program
does not disproportionately affect the distance calculations for other program pairs.
\sstar~\cite{li2025sstar} uses a similar pair-wise comparison approach using an LLM to compare two programs,
whereas \diffcodegen's approach is purely computational, making it much more efficient and scalable.

\subsubsection{Clustering algorithm} \label{sec:clustering_algorithm}
With the pairwise distance matrix $D \in \mathbb{R}^{N \times N}$ where $D_{ij} = d(p_i, p_j)$,
\diffcodegen applies hierarchical agglomerative clustering~\cite{cohenaddad2019hierarchical} to partition the $N$ code candidates into two clusters.
The algorithm starts with each program as its own singleton cluster and iteratively merges the two closest clusters until only two clusters remain.
\diffcodegen uses average-linkage (UPGMA)~\cite{tokuda2022revisiting} to measure inter-cluster distance,
which computes the average distance between all pairs of programs across two clusters:
\begin{equation} \label{eq:linkage}
      D_{\text{avg}}(C_a, C_b) = \frac{1}{|C_a| \cdot |C_b|} \sum_{p_i \in C_a} \sum_{p_j \in C_b} d(p_i, p_j)
\end{equation}
Average-linkage is chosen for its robustness to outliers compared to single-linkage,
and its tendency to produce more balanced clusters compared to complete-linkage~\cite{tokuda2022revisiting}.

\subsubsection{Selection strategy}
After clustering, \diffcodegen identifies the larger cluster as the \emph{consensus cluster},
hypothesizing that programs with consistent behavior are more likely to be correct implementations.
From the consensus cluster $C^*$, \diffcodegen selects the \emph{medoid},
which is the program with the minimum total behavioral distance to all other programs within the same cluster:
\begin{equation} \label{eq:medoid}
      p^* = \argmin_{p_i \in C^*} \sum_{p_j \in C^*, j \neq i} d(p_i, p_j)
\end{equation}
The medoid represents the most central program in the behavioral space of the consensus cluster,
making it a robust choice that best represents the cluster's collective behavior.
If the two clusters have equal sizes, \diffcodegen arbitrarily selects one as the consensus cluster.
If only one program is generated or all programs exhibit identical behavior,
\diffcodegen chooses that program directly.

\section{Evaluation}

We evaluate the performance of \diffcodegen{} from three aspects: the performance gain from different differential methods,
the comparison with state-of-the-art LLM-based test-time scaling methods,
and the cost and overhead analysis.

\paragraph{Experimental setup}
For improvement analysis,
we evaluate on recent state-of-the-art foundation models,
including GPT-4.1, GPT-4o, GPT-5, and coding specific model Qwen2.5\-Coder\-Instruct~\cite{hui2024qwen25coder}.
For comparison with test-time scaling methods,
we follow \sstar~\cite{li2025sstar} to select Qwen2.5\-Coder~\cite{hui2024qwen25coder}
and GPT-4o-mini to keep a fair comparison on the same benchmark \lcb~\cite{jain2025livecodebench}.

\paragraph{Implementation details}
We use vLLM~\cite{kwon2023efficient} to serve open-weight models,
and use the official APIs for closed-weight models.
We implement \diffcodegen in four stages as described in \autoref{sec:method}:
generation, fuzz, differential, and clustering.
Each stage except clustering is parallelized:
\begin{itemize*}
	\item Generation and differential analysis are IO-bound stages implemented as asynchronous coroutines.
	      For repeated sampling, we directly use the API parameter $n$ if it is supported by the model API.
	      Otherwise, we run $N$ asynchronous requests with the same prompt.
	      For differential prompting (\autoref{sec:diff-prompting}),
	      we use batched prompting, sending all prompts together to vLLM.

	\item We implement the fuzzing stage using multiprocessing to fully utilize CPU resources.
	      We fuzz each candidate program in a separate process for one minute following previous work~\cite{zhao2023understanding,huang2024code,he2025fuzzaug}.
\end{itemize*}
We run Qwen2.5-Coder-32B on 2 NVIDIA H100-80G GPUs with tensor parallelism,
and all other models on a single NVIDIA A100-80G GPU.
We run all CPU-tasks (fuzzing, differential analysis, and clustering) on a machine with
2 AMD EPYC 9354 32-Core CPUs (128 threads in total), 1.48TB RAM, and Ubuntu 22.04 OS.
As with \sstar{}, all experiments are conducted once.

\subsection{Performance gain from different differential methods} \label{sec:gain}

\begin{table}[t]
	\centering
	\caption{%
		Performance gain from \diffcodegen{} with different differential generation methods.
		Rep-N: repeated sampling (\autoref{sec:diff-sampling});
		Beam search: beam search sampling (\autoref{sec:diff-decoding}), `-': not supported by API-access;
		PromptPerturb: differential prompting (\autoref{sec:diff-prompting}).
		Improvements are shown in absolute percentage gain.
	}
	\label{tab:improve}
	\begin{tabular}{clcccc}
    \toprule
    Model                                    & Size & Baseline & Rep-N                 & Beam search          & PromptPerturb        \\
    \midrule
    \rowcolor{gray!20}\multicolumn{6}{c}{\textit{LiveCodeBench release-v2}}                                                          \\
    \multirow{2}{*}{GPT-4.1}                 & nano & 47.4     & 54.4 \emph{(+7.0\%)}  & -                    & 53.6 \emph{(+6.2\%)} \\
                                             & mini & 64.6     & 73.2 \emph{(+8.6\%)}  & -                    & 73.2 \emph{(+8.6\%)} \\
    GPT-5                                    & nano & 88.5     & 91.2 \emph{(+2.7\%)}  & -                    & 91.4 \emph{(+2.9\%)} \\
    GPT-4o                                   & mini & 40.9     & 54.4 \emph{(+13.5\%)} & -                    & 46.4 \emph{(+5.5\%)} \\
    \lightmidrule
    \multirow{2}{*}{Qwen-2.5-Coder-Instruct} & 7B   & 29.4     & 38.9 \emph{(+9.5\%)}  & 36.6 \emph{(+7.2\%)} & 36.8 \emph{(+7.4\%)} \\
                                             & 14B  & 44.8     & 47.9 \emph{(+3.1\%)}  & 46.6 \emph{(+1.8\%)} & 49.1 \emph{(+4.3\%)} \\
    \midrule
    \rowcolor{gray!20}\multicolumn{6}{c}{\textit{LiveCodeBench release-v5}}                                                          \\
    \multirow{2}{*}{GPT-4.1}                 & nano & 44.6     & 46.1 \emph{(+1.5\%)}  & -                    & 45.1 \emph{(+0.5\%)} \\
                                             & mini & 59.8     & 65.5 \emph{(+5.7\%)}  & -                    & 65.1 \emph{(+5.3\%)} \\
    GPT-5                                    & nano & 82.5     & 87.1 \emph{(+4.6\%)}  & -                    & 85.7 \emph{(+3.2\%)} \\
    \bottomrule
\end{tabular}

\end{table}

We present the performance gain from \diffcodegen{} in \autoref{tab:improve}.
In this experiment,
we set the number of samples $N=18$ for all differential methods.
We conduct our experiments on two non-overlapping versions of \lcb,
release-v2 and release-v5,
to avoid potential overfitting to a single benchmark.
The results indicate that \diffcodegen{} consistently improves the base models' performance
across all evaluated models and all differential methods.
The improvements range from +0.5\% to +13.5\% absolute gain,
with models at lower baselines generally showing larger improvements due to ceiling effects.
While accuracy gains vary across settings,
the key advantage of \diffcodegen{} lies in its efficiency:
it achieves these improvements at a fraction of the cost of LLM-based selection methods
(see \autoref{fig:exec_time} and \autoref{fig:token_usage}).
Beam search is only available for open-weight models served locally,
as closed-weight APIs do not expose decoding-level control.
Despite this limitation, beam search still provides consistent improvements over the baseline,
although not as pronounced as repeated sampling.
We include all differential methods in \diffcodegen{} for completeness,
allowing practitioners to choose based on their deployment constraints.

\subsubsection{Multi-round evaluation} \label{sec:multi-eval}
\begin{wraptable}{r}{0.5\columnwidth}
	\centering
	\caption{Multi-round evaluation of \diffcodegen{}.}
	\label{tab:multi-eval}
	\small
\begin{tabular}{clccc}
    \toprule
    Model                    & Size                  & Round & Rep-N & PromptPerturb \\
    \midrule
    \multirow{6}{*}{GPT-4.1} & \multirow{3}{*}{nano} & 1     & 46.14 & 45.11         \\
                             &                       & 2     & 46.14 & 44.43         \\
                             &                       & 3     & 46.70 & 44.65         \\
    \lightcmidrule{3-5}
                             & \multirow{3}{*}{mini} & 1     & 65.45 & 65.11         \\
                             &                       & 2     & 65.80 & 63.41         \\
                             &                       & 3     & 64.89 & 63.52         \\
    \midrule
    \multirow{3}{*}{GPT-5}   & \multirow{3}{*}{nano} & 1     & 87.05 & 85.68         \\
                             &                       & 2     & 86.48 & 85.57         \\
                             &                       & 3     & 86.70 & 86.36         \\
    \bottomrule
\end{tabular}

\end{wraptable}

To further demonstrate the stability of \diffcodegen{} for evaluation rigor,
we run the experiments on \lcb release-v5 three times independently.
We present the results in \autoref{tab:multi-eval}.
Across all configurations,
the results exhibit low variance across runs,
with a coefficient of variation (CV) below 1.5\% in every case.
Specifically,
repeated sampling yields a maximum standard deviation of 0.46
and a maximum range of 0.91 percentage points across three runs,
while differential prompting shows a maximum standard deviation of 0.95
and a maximum range of 1.70 percentage points.
Repeated sampling is consistently more stable than differential prompting
(mean CV 0.58\% vs.\ 0.92\%)
and also consistently achieves higher accuracy,
outperforming differential prompting
by 0.9--1.6 percentage points in mean \passatone{} across all configurations.

\subsection{Comparing with test-time scaling methods} \label{sec:comparison-sota}

\subsubsection{Baselines}
We evaluate \diffcodegen{} against two state-of-the-art LLM-based test-time scaling methods:
majority voting~\cite{li2022alphacode,wang2023selfconsistency} and SkyCoder~\cite{li2025sstar}.
Majority voting first generates $N$ samples via explicit repeated sampling,
then uses an LLM to generate test cases for each candidate program,
and selects the program with the maximum agreement score on the generated test cases.
SkyCoder is a variant of \sstar{} with the requirement of using existing public tests removed.
SkyCoder utilizes \emph{adaptive input synthesis} to select the best program among $N$ candidates.
The selection is based on pairwise agreement scoring:
for each pair of candidates, SkyCoder compares their outputs on each generated test case,
and increments both candidates' scores if their outputs match.
The candidate with the highest total agreement score is selected as the final output.
We directly adopt the implementation from the official \sstar{} release~\cite{Sstar_github}.

\subsubsection{Comparison results}
For comparison evaluation,
we set all hyperparameters to 18 parallel samples ($N=18$)
and disable access to public tests for all methods to ensure a fair comparison.
For majority voting and SkyCoder,
we use the same model for their selection stage as their generation stage.
We present the comparison results in \autoref{tab:sota}.
We emphasize that the primary contribution of \diffcodegen{} is its efficiency,
not maximizing accuracy over all baselines.
Nevertheless, \diffcodegen{} outperforms the base model and SkyCoder across all models,
and achieves competitive or superior accuracy compared to majority voting.
Notably, all hyperparameters (\eg, temperature, fuzzing time, number of clusters)
are adopted directly from prior work~\cite{zhao2023understanding,huang2024code,he2025fuzzaug,li2025sstar,shi2024can}
without any specific tuning,
suggesting that further gains may be achievable with dedicated optimization.

\begin{table}[t]
	\centering
	\caption{%
		Comparison with existing LLM-based test-time scaling methods.
		The best results are in \textbf{bold},
		and the second best are \underline{underlined}.
	}
	\label{tab:sota}
	\begin{tabular}{lccccccc}
    \toprule
                        & \multicolumn{6}{c}{Qwen-2.5-Coder-Instruct} & 4o-mini                                                                                                         \\
    \cmidrule(lr){2-7}
    Method              & 0.5B                                        & 1.5B             & 3B               & 7B               & 14B              & 32B              &                  \\
    \midrule
    Zero-shot           & 1.2                                         & 7.0              & 7.0              & 29.4             & 44.8             & \underline{47.4} & 40.9             \\
    Majority Vote       & 2.5                                         & 11.0             & \underline{25.9} & \underline{37.5} & \textbf{50.8}    & 45.5             & 46.6             \\
    SkyCoder            & \underline{2.9}                             & \underline{12.5} & 22.6             & 35.6             & 46.1             & 44.2             & \underline{47.7} \\
    DiffCodeGen (Rep-N) & \textbf{5.1}                                & \textbf{13.5}    & \textbf{26.2}    & \textbf{38.94}   & \underline{47.9} & \textbf{54.0}    & \textbf{54.4}    \\
    \bottomrule
\end{tabular}

\end{table}

\subsection{Token cost and overhead analysis}

To demonstrate where \diffcodegen really shines,
we analyze its cost and overhead compared to existing LLM-based test-time scaling methods.
We evaluate this from two end-to-end aspects:
total token usage and total execution time.
We select two representative models,
Qwen2.5-Coder-7B-Instruct with vLLM and GPT-4o-mini with OpenAI API,
to cover both the self-hosted compute-limited scenario and the API-based scenario.
We conduct experiments on the same machine with one NVIDIA A100-80G GPU and two AMD EPYC 9354 CPUs.

\begin{figure}
	\includegraphics[width=\textwidth]{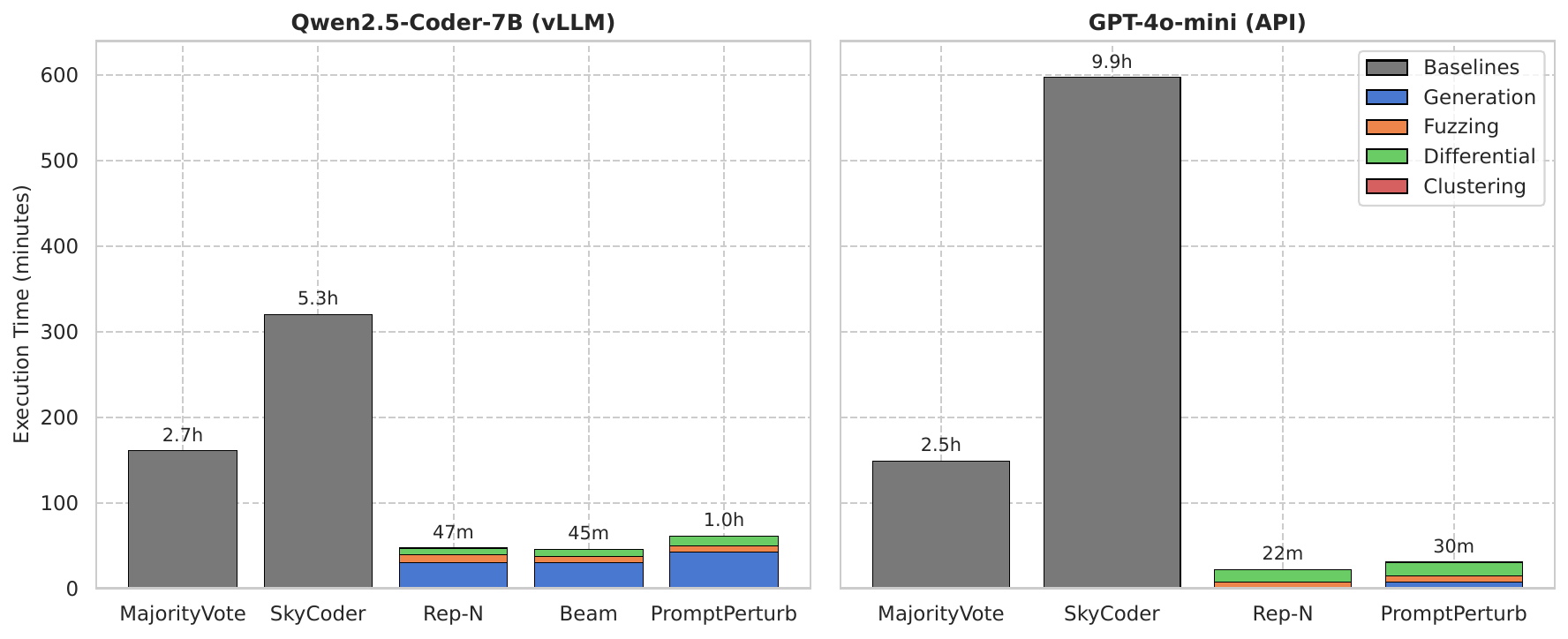}
	\caption{Execution time comparison among different test-time scaling methods.}
	\label{fig:exec_time}
	\Description{Execution time comparison.}
\end{figure}

\subsubsection{Execution time analysis}
The execution time comparison is presented in \autoref{fig:exec_time}.
Compared to majority voting and SkyCoder,
\diffcodegen significantly reduces total execution time across both deployment scenarios.
All variants of \diffcodegen take only one-fifth of the total execution time of majority voting on locally-served models
(47m for Rep-N vs. 2.7h for majority voting),
and achieve approximately 5\% of the time compared to SkyCoder on API-based models (22m vs. 9.9h).
Examining the time breakdown reveals where \diffcodegen's efficiency originates.
For \diffcodegen variants, fuzzing dominates the runtime, while generation, differential analysis, and clustering contribute minimally. Notably, clustering overhead is negligible,
barely visible in the figure.
This confirms that our distance-based clustering approach (\autoref{sec:selection}) adds virtually no computational cost compared to LLM-based selection methods.
Among \diffcodegen variants, beam search is marginally faster than repeated sampling on the local model.
However, beam search is not available for API-based models, limiting its applicability and our potential for further analysis.
Prompt perturbation takes slightly longer (1.0h on vLLM, 30m on API) due to the overhead of processing 18 distinct prompts,
though it remains substantially faster than the baselines.

\begin{figure}
	\includegraphics[width=\textwidth]{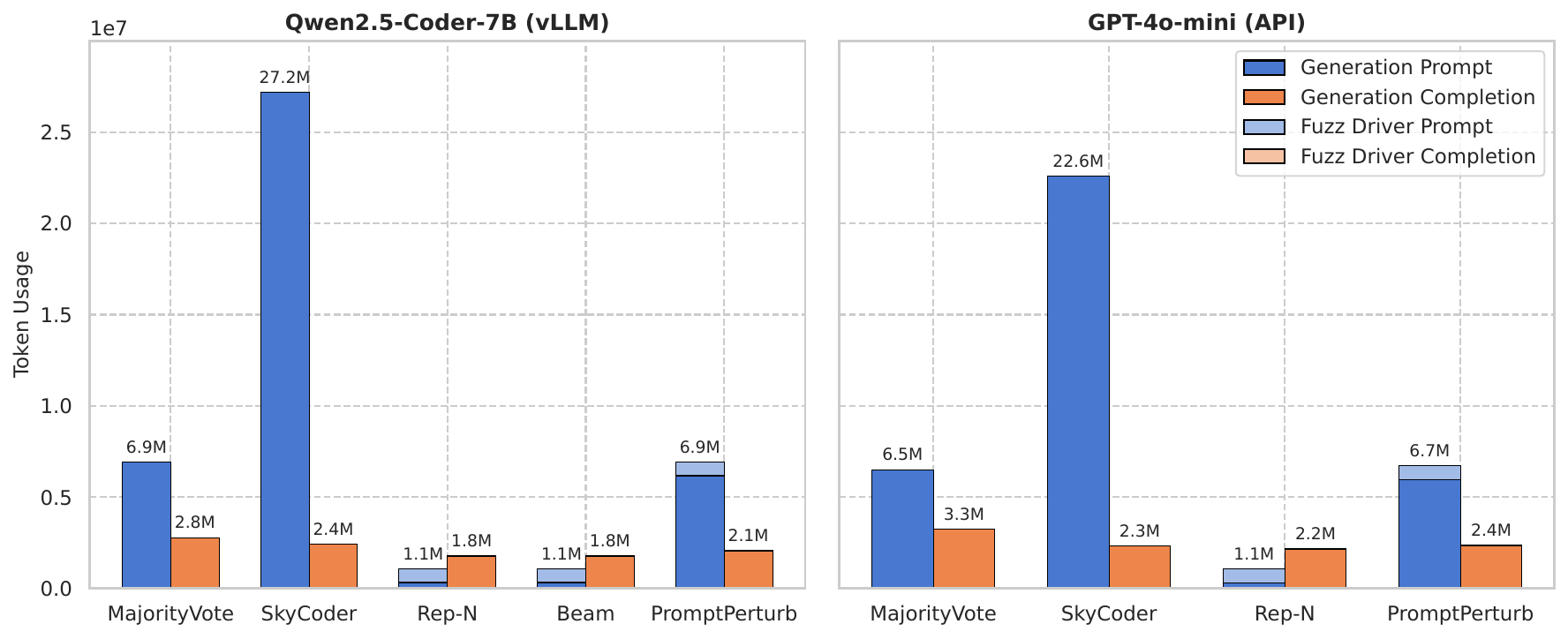}
	\caption{Token usage comparison among different test-time scaling methods. `Prompt` uses input tokens, and `completion` uses output tokens.}
	\label{fig:token_usage}
	\Description{Token usage comparison.}
\end{figure}

\subsubsection{Token usage analysis}
The token usage comparison is presented in \autoref{fig:token_usage}.
The primary advantage of \diffcodegen stems from the dramatic reduction in input tokens consumed during the input generation phase and selection phase.
Majority voting and SkyCoder both require feeding generated code candidates back into the LLM for test synthesis and pairwise comparison,
resulting in 6.9M and 27.2M input tokens respectively on Qwen2.5-Coder-7B.
In contrast, \diffcodegen with repeated sampling consumes only 1.1M input tokens,
approximately \emph{4\%} of SkyCoder's consumption,
since it performs selection via differential analysis rather than additional LLM inference.
For \diffcodegen variants, prompt perturbation uses the highest token cost (6.9M input tokens on Qwen, 6.7M on GPT-4o-mini),
which is at the same level as majority voting.
This is expected since each of the 18 perturbations produces a distinct prompt that must be processed separately,
whereas repeated sampling and beam search reuse the same prompt
and benefit from prefix caching in modern LLM serving frameworks~\cite{kwon2023efficient}.
With prefix caching, the shared prompt is processed once and reused across all $N$ samples,
resulting in only 1.1M input tokens.

\diffcodegen requires a small additional token cost to generate a fuzz driver in library function mode (\autoref{sec:fuzzing}).
As shown in \autoref{fig:token_usage},
the fuzz driver prompt and completion tokens (visible as small segments in the \diffcodegen bars)
contribute minimally to the total: approximately 0.3M tokens combined.
This cost is shared across all candidates and is negligible compared to the selection overhead of LLM-based methods.
For output tokens,
\diffcodegen with repeated sampling saves approximately 0.6M tokens on Qwen and 0.1M tokens on GPT-4o-mini compared to SkyCoder,
and 1M tokens on Qwen and 1.1M tokens on GPT-4o-mini compared to majority voting.
Overall, \diffcodegen's cost profile makes it particularly attractive for API-based deployment,
where token usage directly translates to monetary cost, and for resource-constrained scenarios where inference budget is the bottleneck.
By performing selection without any additional LLM inference,
\diffcodegen achieves competitive or superior Pass@1 performance (\autoref{sec:comparison-sota})
while consuming a fraction of the resources.

\subsection{Ablation studies}

In this section,
we analyze the key design choices of \diffcodegen{} through ablation studies.
We use GPT-4.1-nano for all experiments in this section
since it is the fastest and most affordable model among all evaluated models,
while providing an official API for repeated sampling.

\subsubsection{Choosing the number of samples}

In \autoref{sec:gain},
we set the number of samples $N=18$ for all differential methods
since we included 18 prompt perturbations in differential prompting.
We keep the same number to study the performance gain from different differential methods.
In this section,
we vary $N$ for the repeated sampling method
to study how \diffcodegen{} scales with the number of samples.
We evaluate $N$ from $\{1, 4, 8, 18\}$.

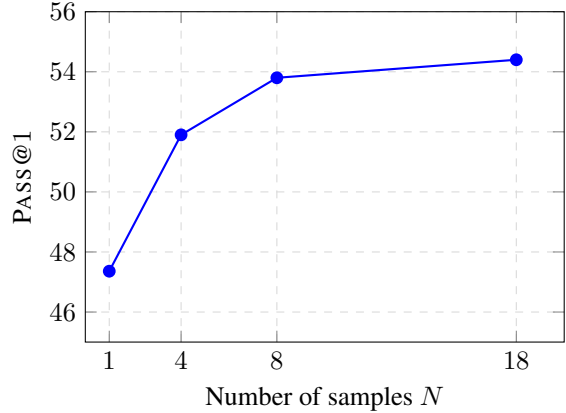
\begin{wrapfigure}{r}{0.5\columnwidth}
	\centering
	\begin{tikzpicture}
    \begin{axis}[
            xlabel={Number of samples $N$},
            ylabel={\passatone},
            xmin=0, xmax=20,
            ymin=45, ymax=56,
            xtick={1,4,8,18},
            ytick={46,48,50,52,54,56},
            legend pos=south east,
            grid=major,
            grid style={dashed,gray!30},
            width=0.48\columnwidth,
            height=0.36\columnwidth,
            mark size=2pt,
        ]
        \addplot[
            color=blue,
            mark=*,
            thick,
        ]
        coordinates {
                (1, 47.36)
                (4, 51.9)
                (8, 53.8)
                (18, 54.40)
            };
    \end{axis}
\end{tikzpicture}
	\caption{Performance scaling with number of samples.}
	\label{fig:ablation-n}
\end{wrapfigure}
Our results are shown in \autoref{fig:ablation-n}.
We observe that \diffcodegen{} consistently improves \passatone as $N$ increases,
demonstrating effective test-time scaling.
The improvement is most pronounced when increasing from $N=1$ (47.36\%) to $N=4$ (51.9\%),
yielding a gain of 4.54\% absolute increase.
Further increases show diminishing returns:
$N=8$ achieves 53.8\% (+1.9\% over $N=4$),
and $N=18$ reaches 54.4\% (+0.6\% over $N=8$).
This diminishing returns pattern is expected,
as additional samples are more likely to produce redundant solutions
rather than exploring new regions of the solution space.
Nevertheless, even modest increases in $N$ provide meaningful improvements,
making \diffcodegen{} a practical choice for scenarios
where additional compute budget is available.
We did not further scale $N$ beyond 18 due to the already existing diminishing returns.
Larger $N$ is also typically not recommended by model providers due to potential degradation in sample quality
(for example, Gemini only supports up to $N=8$~\cite{gemini-doc}).

\subsubsection{Choosing the number of clusters} %
We follow the common practice to cluster the candidate programs for each task into two clusters~\cite{li2025sstar}
in \autoref{sec:gain} and \autoref{sec:comparison-sota}.
However, instead of choosing two clusters ad hoc,
we investigate the impact of varying the number of clusters on the performance of \diffcodegen{}
both empirically and analytically.
To conduct this study,
we vary the number of clusters from 1 to 5,
with all the other hyperparameters kept the same as in \autoref{sec:comparison-sota}.
With one cluster,
all candidate programs are grouped together,
and \autoref{sec:clustering_algorithm} degrades to a pure distance-based selection,
\ie, selecting the medoid among all candidates.

\begin{table}[t]
	\centering
	\caption{Analysis on the number of clusters.
		Silhouette: clustering quality score~\cite{rousseeuw1987silhouettes};
		Largest Ratio: mean $\pm$ std of the largest cluster ratio across tasks;
		Conf: confidence level.
	}
	\label{tab:cluster}
	\begin{tabular}{cccr@{$\,\pm\,$}lcc}
    \toprule
    \# Cluster & \passatone & Silhouette & \multicolumn{2}{c}{Largest Ratio (\%)} & High Conf ($\geq$90\%) & Low Conf ($<$60\%)       \\
    \midrule
    1          & 53.43      & -          & 100.00                                 & 0.00                   & 511                & 0   \\
    2          & 54.40      & 0.5786     & 86.66                                  & 11.86                  & 293                & 22  \\
    3          & 53.03      & 0.4379     & 76.52                                  & 14.72                  & 0                  & 84  \\
    4          & 53.42      & 0.3757     & 69.19                                  & 15.24                  & 0                  & 145 \\
    5          & 53.62      & 0.3174     & 63.04                                  & 15.01                  & 0                  & 205 \\
    \bottomrule
\end{tabular}

\end{table}
\paragraph{Silhouette}
A key question in clustering is whether the partition reflects the genuine structure in the data
or merely imposes unreasonable groupings.
The silhouette coefficient~\cite{rousseeuw1987silhouettes} is a principled measurement of clustering quality.
It compares each object's cohesion within its assigned cluster to its separation from the nearest neighboring cluster,
\ie, the mean intra-cluster distance versus the mean nearest-cluster distance.
Recall from \autoref{eq:distance}, the distance $d(p_i, p_j)$ between two programs $p_i$ and $p_j$
is the fraction of inputs on which they disagree.
Formally, for each object $i$ assigned to cluster $A$, let $a(i)$ denote the average dissimilarity of $i$
to all other objects in $A$:
\[
	a(i) = \frac{1}{|A| - 1} \sum_{j \in A,\, j \neq i} d(p_i, p_j).
\]
Let $b(i) $ denote the smallest average dissimilarity of $i$ to any other cluster $C$:
\[
	b(i) = \min_{C \neq A} \bar{d}(i, C), \\
	\quad\text{where } \bar{d}(i, C) = \frac{1}{|C|} \sum_{j \in C} d(p_i, p_j).
\]
$\bar{d}(i, C)$ is the average distance from $i$ to all members of cluster $C$.
$a(i)$ is the mean within-cluster distance,
and $b(i)$ is the mean nearest-cluster distance.
The silhouette coefficient of object $i$ is:
\begin{equation} \label{eqn:silhouette}
	s(i) = \frac{b(i) - a(i)}{\max\{a(i),\, b(i)\}},
\end{equation}
which is bounded by $-1 \leq s(i) \leq 1$.
A value close to $+1$ indicates that $i$ is well within its cluster and far from the nearest alternative;
a value near $0$ means $i$ lies on the boundary between two clusters;
and a value close to $-1$ suggests $i$ may be misclassified.

\paragraph{Largest cluster ratio and confidence}
While the silhouette coefficient measures geometric separation quality,
it does not directly capture the degree of behavioral consensus among candidates,
which is the signal our selection method (\autoref{alg:clustering}) exploits.
Therefore, we introduce two complementary metrics.
The \emph{largest cluster ratio} quantifies how dominant the majority behavior is.
For a task whose $N$ candidates are grouped into clusters $C_1, \ldots, C_k$,
its largest cluster ratio is
\[
	r = \frac{\max_{1 \leq j \leq k}\, |C_j|}{N}.
\]
A value of $r$ close to $1$ indicates that most candidates exhibit similar behavioral patterns,
suggesting strong agreement on the solution.
We report the mean and standard deviation of $r$ across all tasks.
We further categorize tasks by confidence:
\emph{high confidence} tasks have $r \geq 0.9$
(at least 90\% of candidates in the largest cluster),
while \emph{low confidence} tasks have $r < 0.6$
(less than 60\% in the largest cluster).
These thresholds provide an interpretable summary
of the distribution of consensus levels across the benchmark.

We present our cluster analysis results in \autoref{tab:cluster}.
Our analysis supports $k=2$ as the optimal choice from two perspectives.
\begin{itemize*}
	\item \emph{Empirically}, $k=2$ yields the highest \passatone{} (54.40\%),
	      outperforming both the pure-distance baseline ($k=1$, 53.43\%)
	      and all finer-grained partitions ($k \geq 3$).
	      The improvement of $+$0.97\% over $k=1$ confirms that
	      distinguishing majority from minority behaviors adds value beyond pure medoid selection,
	      while further increasing $k$ degrades selection quality.
	\item \emph{Analytically}, two complementary metrics corroborate this finding.
	      The silhouette score peaks at $k=2$ (0.5786) and decreases monotonically as $k$ increases,
	      indicating that two clusters best reflect the natural structure of candidate behaviors,
	      whereas smaller partitions over-segment the solution space.
	      The largest cluster ratio further confirms strong behavioral consensus at $k=2$.
	      Out of 511 tasks, 293 are high-confidence
	      and only 22 tasks are low-confidence.
	      For $k \geq 3$, no task reaches the high-confidence threshold
	      and low-confidence tasks grow rapidly.
\end{itemize*}
These findings validate our choice of two clusters as not merely a convenient default,
but an empirically and analytically justified decision for our setting.

\subsubsection{Model-based selectors over the differential signal}
\label{sec:learned-selectors}

The differential I/O pairs produced by fuzzing and differential testing
can also feed selectors other than distance-based clustering.
We replace the clustering step with three off-the-shelf model-based selectors
and measure their accuracy and cost.
We evaluate \textit{pointwise} scoring (each candidate rated $1$--$10$ by an LLM judge),
\textit{pairwise} judging with positional debiasing
($3N$ sampled pairs per task, each queried twice with swapped order,
aggregated by win-rate counting with $0.5$ credit for ties or disagreements),
and a public \textit{reward model}
(\texttt{Skywork-Reward-V2-Llama-3.1-8B}~\cite{liu2026skyworkrewardv2}, argmax selection).
For both LLM-based selectors we use \texttt{gpt-4.1-mini} as the judge
across all three generation models;
all other settings ($N=18$, fuzzing time, candidate pool, prompts)
match \autoref{sec:multi-eval}.
We report mean \passatone{} on LiveCodeBench \texttt{release\_v5} over two independent runs.

\begin{wraptable}{r}{0.5\columnwidth}
	\caption{%
		\passatone{} of three model-based selectors applied within the \diffcodegen{}
		differential pipeline on LiveCodeBench \texttt{v5},
		compared against \diffcodegen{}'s distance-based clustering
		(mean over 2 runs)
		Cost rows report per-evaluation total tokens (prompt $+$ completion)
		and wall-clock time; $^*$reward model uses A100 GPU time rather than API tokens.
	}
	\label{tab:learned-selectors}
	\centering

	\begin{adjustbox}{max width=\linewidth}
		\begin{tabular}{lcccc}
    \toprule
    Model        & Clustering & Pointwise & Pairwise & Reward \\
    \midrule
    GPT-4.1-nano & 46.42      & 50.85     & 49.09    & 45.57  \\
    GPT-4.1-mini & 65.35      & 65.35     & 66.19    & 63.70  \\
    GPT-5-nano   & 86.59      & 86.54     & 85.29    & 85.40  \\
    \midrule
    Tokens (M)   & n/a        & 14.3      & 100      & --     \\
    Time (min)   & $<$1       & 6.5       & 15       & 20$^*$ \\
    \bottomrule
\end{tabular}

	\end{adjustbox}
\end{wraptable}

\autoref{tab:learned-selectors} shows that no model-based selector beats clustering on every model.
On the two stronger generation models,
all three selectors fall within roughly $1$~percentage point of clustering
($65.35$ and $86.59$ respectively), within run-to-run variance.
Only on GPT-4.1-nano does pointwise scoring exceed clustering by more than a point
($50.85$ vs.\ $46.42$, $+4.4$~pp);
pairwise and the reward model still trail clustering on at least one model each.
Selector costs, however, differ by an order of magnitude.
Pointwise adds roughly $13$M prompt and $1.3$M completion tokens
plus $\sim 6.5$~minutes of judge latency per evaluation.
Pairwise scales to $\sim 100$M prompt tokens ($\sim 7.6\times$ pointwise)
and $14$--$16$~minutes of latency even with sparse sampling.
The reward model avoids API cost but uses $\sim 20$~minutes of A100 GPU time
and gives the lowest \passatone{} of the three on every generation model.
The differential I/O pairs already carry enough signal to separate candidates,
so we do not need a second LRM call or a large reward model for selection;
distance-based clustering over the same signal matches these selectors at near-zero cost.

\subsubsection{Analysis on fuzz driver generation}
\begin{wraptable}{r}{0.45\columnwidth}
	\vspace{-10pt}
	\centering
	\footnotesize
	\caption{Fuzz driver generation success rate.}
	\label{tab:fuzz-driver-success}
	\begin{tabular}{lllc}
    \toprule
    Ver (total) & Method                        & Model    & Succ (\%) \\
    \midrule
    \multirow{7}{*}{v2 (235)}
                & \multirow{3}{*}{Rep-N}        & 4.1-nano & 94.5      \\
                &                               & 4.1-mini & 96.6      \\
                &                               & 4o-mini  & 93.5      \\
    \cmidrule{2-4}
                & \multirow{2}{*}{Beam}         & Qwen-14B & 93.6      \\
                &                               & Qwen-7B  & 91.9      \\
    \cmidrule{2-4}
                & \multirow{2}{*}{Perturbation} & 4.1-mini & 95.3      \\
                &                               & 4o-mini  & 94.5      \\
    \midrule
    \multirow{5}{*}{v5 (381)}
                & \multirow{3}{*}{Rep-N}        & 4.1-nano & 95.5      \\
                &                               & 4.1-mini & 94.8      \\
                &                               & 5-nano   & 96.1      \\
    \cmidrule{2-4}
                & \multirow{2}{*}{Perturbation} & 4.1-nano & 92.9      \\
                &                               & 4.1-mini & 94.8      \\
    \bottomrule
\end{tabular}

\end{wraptable}

\diffcodegen{} uses LLM-generated fuzz drivers for library functions (\autoref{sec:fuzzing}).
We use \texttt{gpt-4.1-nano} to generate the fuzz driver for all tasks,
since it is the fastest and most affordable model available.
Because it is not stronger than other evaluated models,
the performance gain from \diffcodegen{} cannot be attributed to extra model capacity.
\autoref{tab:fuzz-driver-success} reports the success rate
across our two benchmarks, six models, and all differential methods.
The success rate ranges from 91.9\% to 96.6\%, with an average of 94.5\%,
indicating that modern LLMs can reliably generate compilable and executable fuzz drivers
for the function signatures encountered in competitive programming benchmarks.
For the small fraction of cases where the driver fails to compile or execute (3.4\%--8.1\%),
\diffcodegen{} gracefully falls back to returning the reference solution directly,
ensuring the system always produces an output with minimal impact on overall performance.

\section{Discussions}

\subsection{Threats to validity}

\paragraph{Internal validity}
Our evaluation involves several design choices that may affect results.
We set hyperparameters following established practices in the literature
(\autoref{sec:diff-sampling}, \autoref{sec:fuzzing}).
One minute of coverage-guided fuzzing is sufficient for the fuzzer to explore meaningful program paths and collect enough diverse inputs to characterize a program's dynamic behavior,
especially considering that the candidate programs have limited code size.
This choice balances effectiveness and efficiency,
keeping the overall pipeline practical,
where fuzzing remains the dominant but manageable cost in the end-to-end execution time (\autoref{fig:exec_time}).
For the number of candidates $N$, we conduct ablation studies to analyze its impact on performance.
The selection of the reference solution for fuzzing could influence the quality of generated inputs.
We mitigate this by using principled selection strategies (random for sampling, top-1 for beam search, original prompt for perturbations).
Moreover, since all candidates are generated from the same natural language specification,
implementations should express identical semantics,
so the choice of reference should not have a significant impact.

\paragraph{External validity}
We evaluate \diffcodegen on two non-overlapping versions of \lcb (v2 and v5),
which cover distinct time periods of competitive programming problems and reduce the risk of benchmark-specific overfitting.
The benchmark includes both script-based problems (requiring complete programs with I/O handling) and library function-based problems (requiring standalone functions),
covering different code generation paradigms.
While \lcb provides diverse algorithmic problems,
it may not fully represent real-world code generation scenarios involving complex dependencies, external APIs, or domain-specific libraries.
We evaluate across a range of LLMs from different families and scales,
but results may differ for models with substantially different architectures or training paradigms.

\subsection{Limitations and future work}

\paragraph{Supported programming languages}
\diffcodegen currently supports Python through \pyafl and \atheris.
We select Python because it is the most studied language in LLM-based code generation~\cite{chen2023codet,shi2024can,li2025sstar}.
Extending to other languages is straightforward since coverage-guided fuzzing infrastructure is well-established for many languages,
including C/C++ (\aflpp, \libfuzzer),
Rust (cargo-fuzz), Go (go-fuzz), and Java (Jazzer).
Prior work has demonstrated the integration of coverage-guided fuzzing with code understanding and generation tasks for languages like Java, C/C++, and Rust~\cite{zhao2023understanding,he2025fuzzaug},
suggesting the extensibility of \diffcodegen to additional languages.

\paragraph{Complex input types}
Coverage-guided fuzzing excels at generating primitive types and simple data structures,
but struggles with complex inputs such as custom objects with invariants,
file handles, network connections, or database cursors.
For functions requiring such inputs, the LLM-generated fuzz driver may fail to produce meaningful test cases,
causing \diffcodegen to fall back to the reference solution without differential analysis.
Future work could explore property-based testing~\cite{claessen2000quick_check},
which uses generators to produce well-formed instances of complex data types.
Property-based fuzzing and differential testing are often used in practical software testing~\cite{goldstein2024pbt},
offering a promising direction for extending \diffcodegen to handle such inputs.
\llmcodechoice~\cite{fan2024oracle} is a notable work in this direction;
however, it relies on external human-written oracles and iterative LLM inference.

\paragraph{Non-deterministic programs}
Programs with inherent non-determinism (\eg, involving randomness, timestamps, or concurrency)
may produce different outputs across executions, even when functionally correct.
\diffcodegen's distance metric treats such variations as behavioral differences,
potentially misclassifying correct implementations as outliers.
In this case,
deterministic program analysis techniques are unlikely to help,
whereas embedding-based techniques with dynamic behaviors~\cite{zhao2023understanding,huang2024code} may provide a viable alternative.

\paragraph{Failure of the majority assumption}
\diffcodegen relies on the hypothesis that the largest behavioral cluster contains correct solutions.
This assumption may fail when:
\begin{enumerate*}
    \item most candidates share a common misconception;
    \item the natural language specification is ambiguous, leading to multiple valid interpretations.
\end{enumerate*}
In such cases, \diffcodegen may confidently select an incorrect solution.
However, we have shown empirically that this assumption holds for the vast majority of programming tasks.
Much research has applied differential testing with such assumptions
to complex software systems like compilers and databases~\cite{yang2011finding,chen2016jvm,yang2019hunting,rigger2020sqlancer},

\section{Related work}

\subsection{Code generation models}
Large language models have demonstrated remarkable capabilities in writing programs.
The text-to-code generation task involves LLMs producing code snippets
based on natural language descriptions or specifications~\cite{jiang2026llmcodegen}.
Early work such as Codex~\cite{chen2021humaneval} and AlphaCode~\cite{li2022alphacode} demonstrated that LLMs fine-tuned on code
can solve programming problems described in natural language.
Subsequent models have continued to advance the state of the art,
including open-weight models such as Qwen-Coder~\cite{hui2024qwen25coder,yang2025qwen3}
and reasoning-augmented models like DeepSeek-R1~\cite{Guo2025deepseekr1}.
Generative code models have deeply influenced software development practices,
including resolving issues~\cite{jimenez2024swebench},
automating code reviews~\cite{guo2024exploring},
and repairing programs~\cite{xia2023automated}.
Modern coding agents can accomplish complex software engineering tasks
based on the high-level natural language instructions of users~\cite{xia2025demystifying}.
Recent work also applied LLMs to aid software testing and analysis~\cite{he2024unitsyn,lyu2024promptfuzz,xiong2024program,he2025fuzzaug,zhang2026llamafuzz}.

\subsection{Test-time scaling for code generation}
Test-time scaling (TTS) has emerged as a promising direction for improving LLM performance
by allocating additional compute at inference time~\cite{snell2025scaling,saadfalcon2025an}.
TTS techniques have two directions:
\begin{enumerate*}
    \item workflow-based approaches that utilize task-specific tailored multi-step prompting to enhance the output~\cite{wei2022cot,zhang2025selfanchor},
    \item post-training approaches to train the model to reason at test time by themselves~\cite{deepseekai2025deepseekr1,muennighoff2025s1,li2025skyt1}.
\end{enumerate*}
Workflow-based approaches such as chain-of-thought prompting~\cite{wei2022cot} and self-consistency~\cite{wang2023selfconsistency}
improve performance by generating multiple reasoning paths and aggregating their outputs.
These methods can be applied to LLMs in a plug-and-play manner without additional training.
Self-improvement techniques~\cite{madaan2023selfrefine,yao2023react,shinn2023reflexion,huang2023large,wang2024a}
use iterative refinement loops to let the LLM find better solutions based on its own feedback.
DeepSeek-R1~\cite{deepseekai2025deepseekr1} and s1~\cite{muennighoff2025s1}
demonstrate that scaling test-time compute through extended reasoning learned from reinforcement learning
can further incentivize the model to explore the solution space and find the higher-quality final output.

TTS has also been explored for code generation very recently~\cite{ridnik2024alphacodium,huang2024enhancing,wang2024repogenreflex,li2025rethinkmcts}.
\sstar~\cite{li2025sstar} is a workflow-based TTS method that
uses existing public test cases to iteratively debug generated code,
then synthesizes additional inputs via LLMs to select the best candidate with pair-wise comparison.
While effective, this approach assumes access to public tests that are often unavailable in practice,
and requires significant additional LLM inference for input synthesis.
Furthermore, this selection method is highly inefficient with respect to time and computational resources.
Self-debugging with self-generated tests~\cite{chen2025revisit} is another recent approach
to mitigate the test-available assumption by generating tests using LLMs themselves;
however, it still relies on multiple rounds of LLM inference for both test generation and candidate selection.
ACECODER~\cite{zeng2025acecoder} uses reinforcement learning with automated test-case synthesis
to improve code generation, requiring extensive post-training.
In contrast, \diffcodegen operates without requiring existing tests, additional LLM inference, or training,
performing selection through program analysis instead of additional model calls.

\subsection{Program analysis for large language models}
Program analysis techniques have been increasingly applied to enhance LLM-based code understanding and generation.
\textcite{zhao2023understanding} and \textcite{huang2024code} demonstrate that augmenting LLMs with program execution information
obtained through fuzzing-generated test cases improves code understanding tasks.
\fuzzaug~\cite{he2025fuzzaug} uses coverage-guided fuzzing to generate diverse test inputs
for augmenting neural test generation training data.
These approaches share our insight that program execution provides semantic information
complementary to LLMs.

\section{Conclusion}

In this paper, we presented \diffcodegen, a novel test-time scaling approach for code generation.
By integrating coverage-guided differential analysis,
\diffcodegen effectively and efficiently selects high-quality solutions from multiple candidates generated by LLMs.
Our extensive experiments demonstrate that \diffcodegen consistently improves performance across \Nmodels LLMs,
achieving competitive or superior results compared to state-of-the-art test-time scaling methods.
The key advantage of \diffcodegen lies in its efficiency:
by eliminating the additional LLM inference that prior test-time scaling methods require for candidate selection,
\diffcodegen requires only a small fraction of the execution time compared to existing methods, whether using locally-served or API-accessed models,
while consuming less than 4\% of the total tokens compared to existing methods.
\diffcodegen is model-agnostic, fully asynchronous, and can be combined with reasoning models to further boost performance,
making it a practical solution for real-world code generation and LLM-based software engineering tasks.
We will open-source our implementation to facilitate future research on leveraging software testing and analysis techniques for AI-based software engineering.

\section*{Data availability}
Our artifacts and source code to reproduce the data are available on public anonymous
digital repository Zenodo~\cite{diffcodegen_artifacts}.
We our source code publicly on GitHub at \url{https://github.com/SecurityLab-UCD/DiffCodeGen}
to facilitate future research on software analysis to improve LLM-based code generation.

\section*{Acknowledgments}
This material is based upon work supported by UC Noyce Initiative.

\printbibliography

\end{document}